# Spherical harmonics for fractional quantum numbers *l*=1/n

Lv, Qingzhang
(School of Chemistry and Chemical Engineering, Henan Normal University, Henan, P. R. China 453007)

**Abstract**     The angular momentum quantum number $l$ of spherical harmonic $Y_{l,m}(\theta,\varphi)$ based on an associated Legendre polynomial is non-negative integer 0, 1, 2, … etc. and must never be a fraction. But the study in this paper found that the quantum number $l$ corresponding to other series of solutions of the associated Legendre equation should be fractions. This paper not only proposed the spherical harmonics of $Y_{\frac{1}{2},\pm\frac{1}{2}}(\theta,\varphi)$ for $l=1/2$, but also spherical harmonics $Y_{\frac{1}{n},\pm\frac{1}{n}}(\theta,\varphi)$ for $l=1/n=1/3, 1/4, 1/5…$, etc. In addition to the spin s=1/2 of electron-like particles, the fractional spin such as 1/3, 1/4, 1/5… were boldly speculated to be verified in this paper. Setting the spin of a particle with only two spin components, up and down, to s=1/2 is not necessarily correct. Based on the symmetry of the plots of $Y_{\frac{1}{n},\pm\frac{1}{n}}(\theta,\varphi)$, three different spin classes of particles are predicted. The first class of particles (s=1/2, 1/6…), resembles electrons, particles with the parallel spin tend to move away from each other, and the second class of particles (s=1/3,1/5…) does not repel each other regardless of whether their spins are parallel or not, and the third class of particles (s=1/4,1/8…) always repel each other and tend to move away regardless of whether their spins are parallel or not. The use of fractional spin can be well illustrated for electrons and protons as reported in the literature. This view may be important for the study of quantum mechanics and elementary particles.

## Introduction

The 2022 Nobel Prize in physics was awarded to three scientists of quantum angular momentum entanglement.[1] The quantized angular momentum is important property of microscopic particle systems. Recently, quantum computing and quantum information technology have become research hotspots around the world. In fact, these developing new quantum technologies are all based on the basic principles of quantum mechanics. There are also problems of quantum mechanics requiring theoretical breakthroughs by now. Spin properties are often given important missions in the study of these quantum technologies. But the spin behavior of microscopic particles has not been specifically described in quantum mechanics by now. For example, the spin state of electrons can only be described simply upward or downward.

Since Planck first proposed quantum theory based on blackbody radiation in the beginning of the 20th century, Einstein proposed the photonic theory based on Planck's theory, Bohr went on to propose the theory of the structure of hydrogen atom, And De Broglie boldly proposed the concept of matter wave based on Einstein's photonic theory. After Shrödinger established the wave dynamics theory of quantum mechanics based on electromagnetic field theory, the theory of quantum mechanics become gradually completed. In addition, there were other great scientists who had made outstanding contributions to the theoretical system of quantum mechanics, such as Heisenberg, Dirac, Born. In 1925, Dutch scientists G. Uhlenbeck and S. A. Goudsmit proposed the hypothesis of electron spin. However, electron spin angular momentum has not had a specific mathematically analytic eigenfunction. At the same time, there are no fractional quantum numbers other than the semi-integer 1/2 quantum number of the electron spin mainly because the existence of fractional quantum numbers has not been proven in practise.

Spherical harmonics $Y_{l,m}(\theta,\varphi)$, the eigenfunctions of the squared angular momentum squared operator $\widehat{L^2}$, are given based on the famous Legendre differential equation, which can only correspond to $l$= 0, 1, 2, 3, …etc. nonnegative integers.[2, 3, 4] There will be no polynomial solution of the Legendre equation corresponding to a semi-integer quantum number such as 1/2 of electronic spin. Spin is intrinsic angular momentum not a classical effect, one cannot hope to understand spin based on classical angular momentum, just analogous to the orbital angular momentum.[5] It is puzzling that the angular momentum of semi-integer quantum numbers has not mathematically resolved eigenfunction in books of quantum chemistry. A polynomial type solution to the Legendre equation corresponding to $l$=1/2 is obviously impossible. However, from a theoretical point of view, there should be spherical harmonic functions for $l$=1/2 since the spin angular momentum is real though the electron rotating has no physical reality. The speculation about the spherical harmonics of electronic spin is verified and confirmed from a famous math web site (DLMF).[6] There are really solutions to the Legendre equation corresponding for $l$=1/2, but they are not polynomials of $\sin\theta$ and (or) $\cos\theta$. At the same time, the possibility of a spherical harmonic function corresponding to $l$=1/n has also been confirmed from the web site. These fractional quantum numbers may have potential physical significance in the field of particle physics.

## Theory of fraction angular momentum quantum numbers

Equation 1 is the eigenequation of squared angular momentum operator $\widehat{L^2}$.

$$-\left(\frac{\hbar}{2\pi}\right)^2\left[\frac{1}{\sin\theta}\frac{\partial}{\partial\theta}\left(\sin\theta\frac{\partial}{\partial\theta}\right)+\frac{1}{\sin^2\theta}\frac{\partial^2}{\partial^2\varphi}\right]Y(\theta,\varphi)=k\left(\frac{\hbar}{2\pi}\right)^2 Y(\theta,\varphi) \qquad (1)$$

The eigenfunctions of $\widehat{L^2}$ are the spherical harmonic functions $Y_{l,m}(\theta,\varphi)$. This equation continues to separate $\theta$ and $\varphi$ parts will obtain the following two eigenequations including only one variable separately as below.

$$\frac{1}{\sin\theta}\frac{\partial}{\partial\theta}\left(\sin\theta\frac{\partial\Theta(\theta)}{\partial\theta}\right)-\frac{m^2}{\sin^2\theta}\Theta(\theta)+k\Theta(\theta)=0 \qquad (2)$$

$$\frac{\partial^2\Phi(\varphi)}{\partial^2\varphi}+m^2\Phi(\varphi)=0 \qquad (3)$$

Equation 2 namely $\Theta(\theta)$ equation becomes the standard famous associated Legendre equation when $\cos\theta$ is used as its independent variable, as equation 4.

$$\frac{\partial}{\partial x}\left[(1-x^2)\frac{\partial y(x)}{\partial x}\right] - \frac{m^2}{1-x^2}y(x) + ky(x) = 0 \qquad (4)$$

According to the principle of quantum mechanics, the eigenfunction of the Hermitian operator must be single-valued and continuous in the 3D coordinate space, and so the m quantum number must be an integer like 0, ±1, ±2, …. Only if $k=l(l+1)$, $l=0, 1, 2, 3, …$ etc. nonnegative integers and m = 0, ±1, ±2, … ,±l there will be convergent solutions for the equation, which are a set of associated Legendre polynomials $P_l^m(x)$.

$$y(x) = P_l^m(x) = (1-x^2)^{\frac{m}{2}}\frac{d^m}{dx^m}P_l(x) \qquad \text{(associated Legendre polynomials)}$$

$$P_l(x) = \frac{1}{2^l l!}\frac{d^l}{dx^l}[(x^2-1)^l] \qquad \text{(Legendre polynomials)}$$

So we get the eigenfunctions $Y_{l,m}(\theta,\varphi)$ of equation 1.

$$Y_{l,m}(\theta,\varphi) = N \cdot P_l^m(\cos\theta) \cdot e^{im\varphi}$$

However, the fact that the spin angular momentum quantum number of electron is $s = \frac{1}{2}$ and spin magnetic quantum number is $m_s = \pm\frac{1}{2}$ is reasonably and logically inferred based on experimental facts. Because it is impossible to understand the internal structure of electrons and even if there are spherical harmonics corresponding to $l=1/2$, the $\Phi_m(\varphi)$ function necessarily does not follow the multi-valued prohibition of the wave function, and quantum mechanics takes the spin of electrons as an assumption. And all spin angular momentum theories are expressed by analogy to the orbital angular momentum theory. However, the electron spin quantum number $s=\frac{1}{2}$ exists and should have had its own eigenfunction.

According to the characteristics of the eigenfunctions of Equation 2, each $\Theta_{l,m}(\theta) = CP_l^m(\cos\theta)$ is a polynomial about $\sin\theta$ and $\cos\theta$, and the maximum value of the sum of powers of $\sin\theta$ and $\cos\theta$ corresponds to the value of $l$.[2,3] It can be inferred that the eigenfunction corresponding to $l=\frac{1}{2}$ should have the form of $(\sin\theta)^{1/2}$ but $(\cos\theta)^{1/2}$ since the value of $\cos\theta$ can be negative when the value of $\theta$ varies in the range of 0 ~ π. And the value of m corresponding to $l=\frac{1}{2}$ should be $m=\pm\frac{1}{2}$. The function $\Theta(\theta)=A(\sin\theta)^{1/2}$ was verified by Equation (2) and conformed in § 14.4 of DLMF web site. The solution of Equation (2) for $l=\frac{1}{2}$, $m=\pm\frac{1}{2}$ is:

$$\Theta_{\frac{1}{2},\pm\frac{1}{2}}(\theta) = \sqrt{\frac{2}{\pi}\sin\theta}$$

Combining it with the corresponding the eigenfunction $\Phi_{\pm\frac{1}{2}}(\varphi) = e^{\pm i\frac{\varphi}{2}}$ of equation (3) yields the spherical harmonics $Y_{\frac{1}{2},\pm\frac{1}{2}}(\theta,\varphi)$.

$$Y_{\frac{1}{2},\pm\frac{1}{2}}(\theta,\varphi) = \Theta_{\frac{1}{2},\pm\frac{1}{2}}(\theta) \cdot \Phi_{\pm\frac{1}{2}}(\varphi) = N\cdot\sqrt{\frac{2}{\pi}\sin\theta}\cdot e^{\pm i\frac{\varphi}{2}}$$

Here, the period of $\varphi$ angle is $4\pi$ and $\Phi_{\pm\frac{1}{2}}(\varphi)$ have two values at each coordinate position according to two angle values of $\varphi$ and $2\pi + \varphi$. So the electron spin sphere harmonics break the multi-valued prohibition of the wave function in quantum mechanics. This result encouraged me to consider the probability of $\Theta_{\frac{1}{n}, \pm\frac{1}{n}}(\theta)$ for $l=1/n$ because similar cases with period of $2n\pi$ (here n is an integer) also have certain degree of rationality.

Put $k=l(l+1)$, $l=\frac{1}{n}$ and $m=\pm\frac{1}{n}$ into the $\Theta(\theta)$ equation and get the following equation 5.

$$\frac{1}{\sin\theta}\frac{\partial}{\partial\theta}\left(\sin\theta\frac{\partial\Theta(\theta)}{\partial\theta}\right) - \frac{\left(\pm\frac{1}{n}\right)^2}{\sin^2\theta}\Theta(\theta) + \frac{1}{n}\left(1+\frac{1}{n}\right)\Theta(\theta)=0 \qquad (5)$$

The reasonably speculated solution $\Theta_{\frac{1}{n}, \pm\frac{1}{n}}(\theta) = A(\sin\theta)^{\frac{1}{n}}$ is verified and also be confirmed as Legendre functions $P_\nu^\nu(\cos\theta)$ in DLMF web site (http://dlmf.nist.gov/14.5.E18).

$$P_\nu^\nu(\cos\theta) = \frac{(\sin\theta)^\nu}{2^\nu \Gamma(\nu+1)}$$

$P_\nu^\nu(\cos\theta)$ is Legendre function for real $\nu$. But the values of $\nu$ can not be any real numbers only just $\nu=1/n$ suitable for the spherical harmonics used in quantum mechanics. This will be discussed later on. The $\Theta_{\frac{1}{n}, \pm\frac{1}{n}}(\theta)$ function combining with the eigenfunction $\Phi_{\pm\frac{1}{n}}(\varphi) = \frac{1}{\sqrt{2n\pi}}e^{\pm i\frac{\varphi}{n}}$ of equation (3) gives spherical harmonics $Y_{\frac{1}{n}, \pm\frac{1}{n}}(\theta, \varphi)$.

$$Y_{\frac{1}{n'}\pm\frac{1}{n}}(\theta, \varphi) = \Theta_{\frac{1}{n'}\pm\frac{1}{n}} \cdot \Phi_{\pm\frac{1}{n}}(\varphi) = \frac{(\sin\theta)^{1/n}}{2^{1/n}\Gamma(1/n+1)} \cdot \frac{1}{\sqrt{2n\pi}} e^{\pm i\frac{1}{n}\varphi} = N \cdot (\sin\theta)^{\frac{1}{n}} \cdot e^{\pm i\frac{1}{n}\varphi}$$

Here the period of $\varphi$ angle is $2n\pi$. In quantum chemistry, real form of function is used to describe electronic orbitals. Combining the two complex functions $Y_{\frac{1}{n'}\pm\frac{1}{n}}(\theta, \varphi)$ as below gives two real form spherical harmonics $Y_{\frac{1}{n'},cos}(\theta, \varphi)$ and $Y_{\frac{1}{n'},sin}(\theta, \varphi)$.

$$Y_{\frac{1}{n'},cos}(\theta, \varphi) = \frac{1}{\sqrt{2}}\left(Y_{\frac{1}{n'}+\frac{1}{n}}(\theta, \varphi) + Y_{\frac{1}{n'}-\frac{1}{n}}(\theta, \varphi)\right) = N' \cdot (\sin\theta)^{\frac{1}{n}} \cdot \cos\left(\frac{1}{n}\varphi\right)$$

$$Y_{\frac{1}{n'},sin}(\theta, \varphi) = \frac{1}{i\sqrt{2}}\left(Y_{\frac{1}{n'}+\frac{1}{n}}(\theta, \varphi) - Y_{\frac{1}{n'}-\frac{1}{n}}(\theta, \varphi)\right) = N' \cdot (\sin\theta)^{\frac{1}{n}} \cdot \sin\left(\frac{1}{n}\varphi\right)$$

Clearly, in addition to the spherical harmonics of $l=0, 1, 2, 3$, there are $Y_{\frac{1}{n'}\pm\frac{1}{n}}(\theta, \varphi)$ of $l = \frac{1}{n}$ suitable for $\widehat{L^2}$. So far, no $Y_{\frac{1}{n'}\pm\frac{1}{n}}(\theta, \varphi)$ were mentioned definitely in literature.

# Graphical structure of $Y_{l,m}(\theta, \varphi)$ of $l = 1/n$

To gain insight into the properties of these new spherical harmonics of $l=1/n$, let's analyze their structural diagrams. The surfaces of the atomic orbitals according to $Y_{l,m}(\theta, \varphi)$ of $l= 0, 1, 2$ are well known as shown in Figure 1.

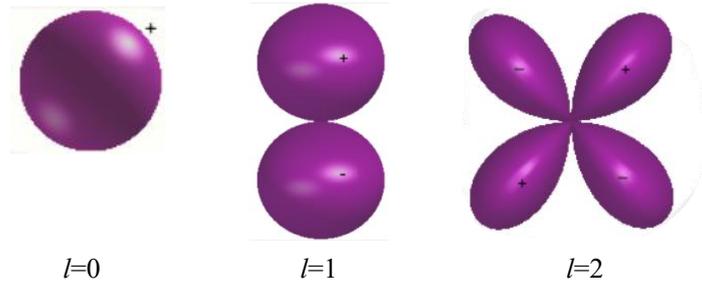

$l=0$          $l=1$          $l=2$

Fig. 1 The structures of real spherical harmonics $Y_{l,m}(\theta,\varphi)$ ($l=0,1,2$).

A common feature of these graphs of real spherical harmonics is that they have $l$ node surfaces clearly between the positive and negative phase regions, so in like manner the surface plots of $Y_{\frac{1}{2},cos}$ and $Y_{\frac{1}{2},sin}$ should have half nodal surface, as shown in Figure 2. The plot of $Y_{\frac{1}{2},sin}$ is symmetrical left and right with that of $Y_{\frac{1}{2},cos}$.

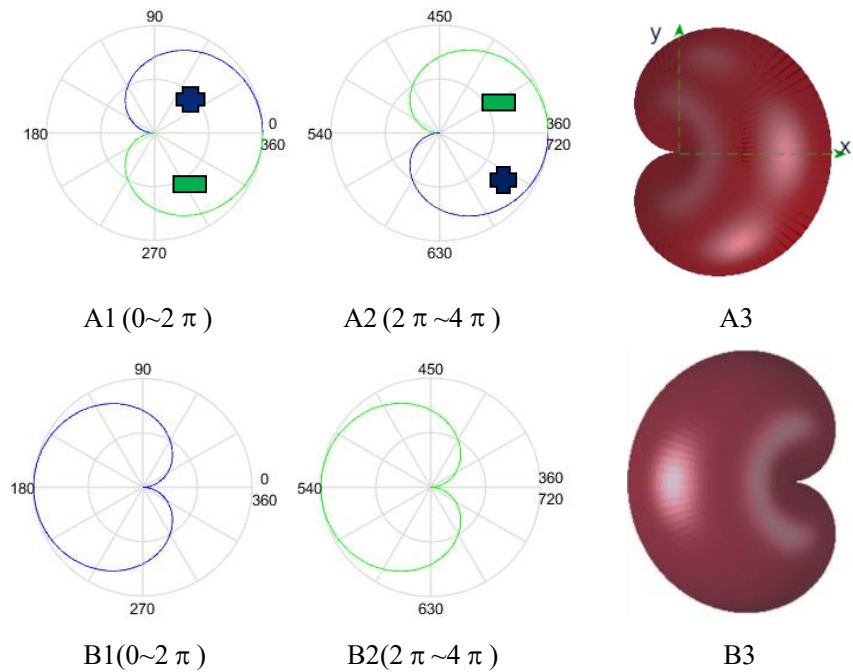

A1 (0~2π)      A2 (2π~4π)      A3

B1 (0~2π)      B2 (2π~4π)      B3

Fig.2 The plots of $Y_{\frac{1}{2},cos}$ (A) and $Y_{\frac{1}{2},sin}$ (B), their positive and negative regions overlap fully(the period of $\varphi$ angle is 4π).

In Figure 2, the positive and negative regions of phase completely overlap. That is, the structure of the positive phase is the same as that of the negative phase. This is different from the atomic orbits shown in Figure 1 with nodes to segment the positive and negative phase regions.

The structure plots corresponding to 1/4 period of the real spherical harmonics $Y_{l,cos}(\theta,\varphi)$ of $l=1/3, 1/4, 1/5$ and $l=1/9$ are presented in Figure 3.

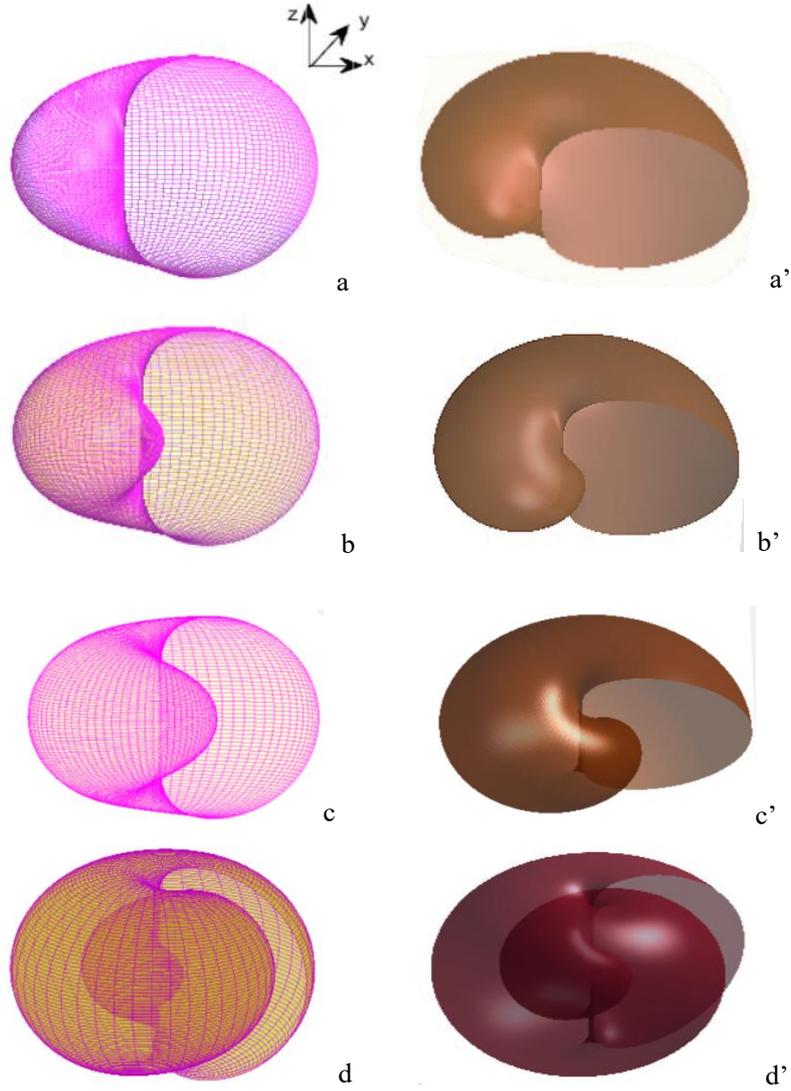

Fig. 3 The plots of $Y_{1/n,\cos}$ in a quarter of period of $\varphi$. (a: $l=\frac{1}{3}$, b: $l=\frac{1}{4}$, c: $l=\frac{1}{5}$ and d: $l=\frac{1}{9}$).

The plots of $Y_{\frac{1}{n},\cos}$, especially the plot of the $Y_{\frac{1}{9},\cos}$, are magical spiral structure as shown in figure 3. Their complete graphs are the repeated overlapping of these plots and their mirror images. Figure 4 is the view of the full period plots in the X-Y plane of $Y_{\frac{1}{3},\cos}$, $Y_{\frac{1}{4},\cos}$ and $Y_{\frac{1}{5},\cos}$. If the rotation axis is Z-axis, then all their full graphics are symmetrical about the X-Y plane, but their symmetry about the X-Z and Y-Z mirrors is not same for different quantum numbers. Since the period of $\varphi$ angle is $2n\pi$ for $Y_{\frac{1}{n},\cos}$, the positive and negative regions overlap certainly. When n is even, its full graphics will be symmetrical with respect to the X-Z plan, and the positive and negative phase regions of $Y_{\frac{1}{n},\cos}$ will completely coincide($l=1/2, 1/4,...$). At the same time, if n=2k (k is odd) the plot of $Y_{\frac{1}{n},\sin}$ is a left and right flip mirror image of $Y_{\frac{1}{n},\cos}$ like in figure 2 ($l=1/2$), and if n=4k (k is an integer) the plot of $Y_{\frac{1}{n},\sin}$ is same of $Y_{\frac{1}{n},\cos}$ ($l=1/4$). When n is odd, the positive and negative phase regions do not overlap exactly, the graph is symmetrical about the X-Z mirror and anti-symmetrical about Y-Z mirror, and the plot of $Y_{\frac{1}{n},sin}$ is the same of $Y_{\frac{1}{n},cos}$ just rotating 90° around z-axis ($l=1/3,1/5...$).

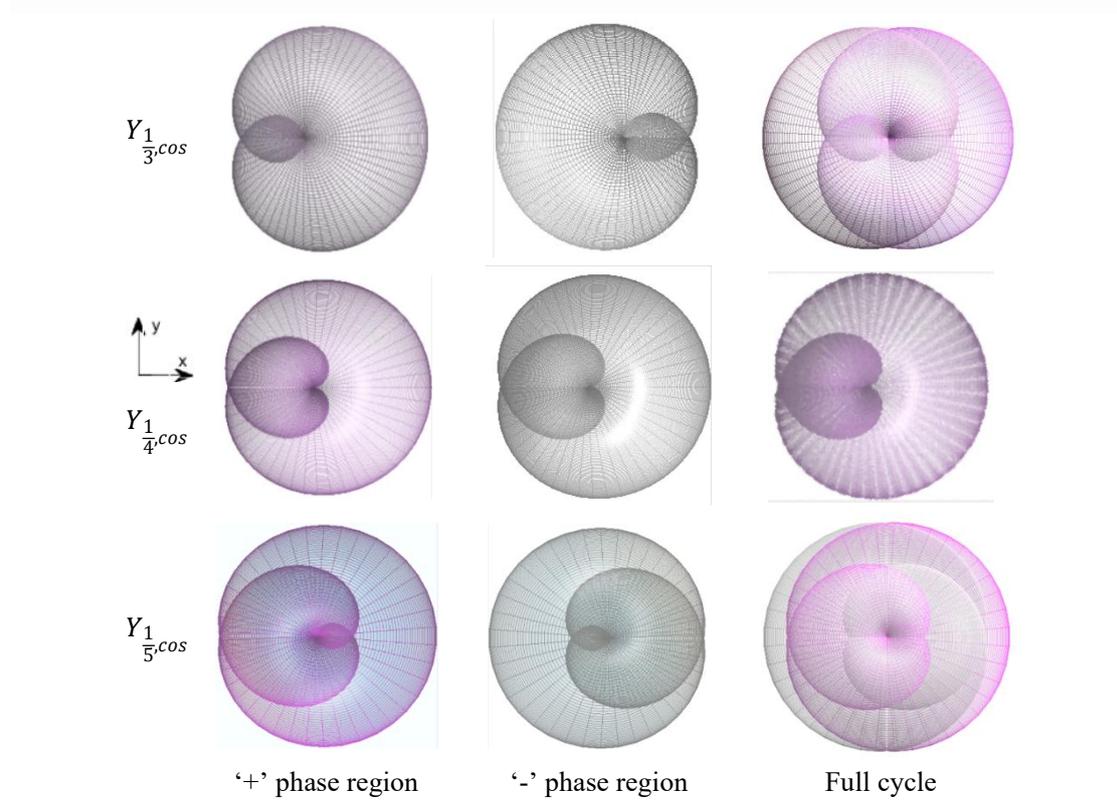

'+' phase region     '−' phase region     Full cycle

Fig. 4 The view in the X-Y plane of spherical harmonics $Y_{\frac{1}{3},cos}$, $Y_{\frac{1}{4},cos}$ and $Y_{\frac{1}{5},cos}$.

## Discussion

In fact the Legendre function $P_\mu^\nu(cos\theta)$ exists for all values of $\nu$ and $\mu$ (DLMF14.3). But only the Legendre polynomials $P_l^m(cos\theta)$ are used to construct the spherical harmonics $Y_{l,m}(\theta,\varphi)$ according the atomic orbitals reported in literature. The reason is that angular momentum and its components are quantized and the spherical harmonics must be single-valued continuous. For s=1/2 of the electron spin, the corresponding spherical harmonic functions break the multivalued prohibition or the angle $\varphi$ has a period of $4\pi$, while it continues to maintain the continuity of the wave functions in the coordinate space. So the periodicity and continuity of the spherical harmonic functions corresponding to the fractional quantum numbers cannot be broken. Then $l$ must take the value of 1/n so that the period of the $\varphi$ angle is $2n\pi$, otherwise the connected two periodic spherical harmonic functions are no longer continuous in coordinate space. For example, the plots of cos type of these Y functions of $l$=2/3, 3/4 and 2/5 are all discontinuous in consecutive two periods although their sin type plots are all continuous as shown in Figure 5. In this case, the pattern of two connected periods, although the same, can not be completely repeated because their initial position is changed, thus forming the precession structure of the complete period pattern, that is, the period is not an integer multiple of $2\pi$. The basic graph of a complete period is the graph structure of the sin shape and the graph of each period has a certain $\varphi$-angle deflection compared with the previous period. If $l$ is an irrational number, its figure will never be able to form a closed graph. That is, the magnitude of the angular momentum is unchanged, but the direction is constantly changed, which violates the conservation principle of the angular momentum.

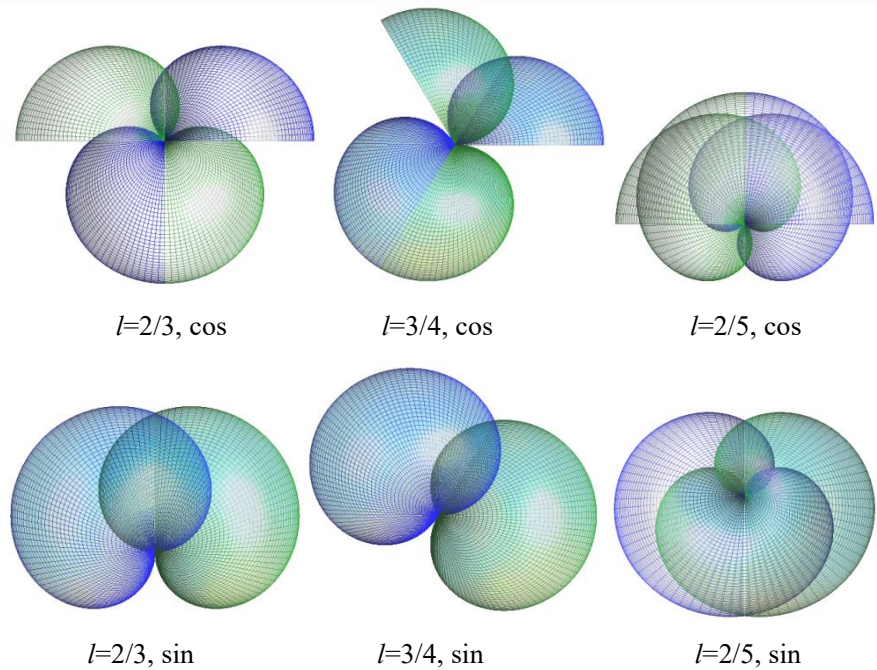

| $l$=2/3, cos | $l$=3/4, cos | $l$=2/5, cos |

| $l$=2/3, sin | $l$=3/4, sin | $l$=2/5, sin |

Fig. 5 The view in the X-Y plane of spherical harmonics for $l$=2/3, 3/4 and 2/5.

Based the theory of atomic orbitals, the values of orbital angular momentum quantum numbers $l$ corresponding to $Y_{l,m}(\theta,\varphi)$ are integer 0, 1, 2, 3, …. Just only the spin angular momentum quantum number of particles such as electrons, protons and neutrons is 1/2. So, one can speculate that may be only the spin quantum numbers of microscopic particles have fractional quantum numbers. So I speculate that there may still be elementary particles in nature with spin quantum numbers of s=1/3, 1/4, 1/5…waiting to be proven, at the same time, the spin component quantum number ms = ± $l$= ±1/n has also two values. So spherical harmonic function for spin can be expressed as $Y_{s,m_s}(\theta,\varphi)$, here $s=\frac{1}{n}$, $m_s=\pm\frac{1}{n}$ and the period of $\varphi$ is 2nπ. The graph structure of these spherical harmonics gradually increases the spherical symmetry as n increases to infinity or $l$=0 until the structure becomes a sphere. Of course, it cannot be ruled out that the orbital motion of some special particles has a fractional angular momentum quantum number, Or the total angular momentum quantum number of the spin and orbital motion of some special particles is fractional. For example, the angular momentum quantum number of the total motion of all particles inside a proton is 1/2. So a fractional spin can also be the total angular momentum quantum number of a particle which we cannot yet observe its internal structure.

Spin properties can be illustrated by $m_s = \pm \frac{1}{n}$ and can be illustrated by spin graphs of real spherical harmonics of *cos* and *sin* types also. The graphical structure of these real spherical harmonics may be related to the spin properties. If the nature of spin is explored based on the graphs of these real spherical harmonics, according to Pauli's theorem, electrons with spin s = 1/2 belongs to fermions, electrons with parallel spins tend to move away, and electrons with antiparallel spins tend to pair. So we can understand it on the symmetry of the plots of *l*=1/2. The spin graphs corresponding to s=1/2, 1/6, … are not centrally symmetric, and spin plots with antiparallel spins are mirror images of each other about Y-Z mirror (as A3 and B3 shown in Figure 2). When particles of s=1/2 with parallel spins approach, their total spin structure asymmetry is exacerbated, and when two particles with antiparallel spins approach the total spin tends to equilibrium, so particles with antiparallel spins have the requirement to be paired, while particles with parallel spins tend to move away. By analogy, particles with spin s=1/3, 1/5, … have central symmetry in their spin plots. When particles with antiparallel spins approach, the spins are more balanced, so that two particles with antiparallel spins tend to be paired, and the repulsion of two particles with parallel spins is not as mutually exclusive as the electrons do or no rejection. Such particles are quasi-bosons. For spins s=1/4, 1/8, 1/12, … particles with antiparallel spins have the same non-central symmetry and the same structure, so such particles may all tend to move away and do not like pairing. Of course, the spherical symmetry of $Y_{\frac{1}{n}, cos}$ plot increases as n increases, so the smaller the spin s= 1/n, the less mutually exclusive the particles, even if their spins are the same.

These fractional quantum numbers and corresponding spherical harmonics may be theoretically useful for the study of particles such as Majorana particles.[7] In addition, fractional spin quantum numbers also indicate that microscopic particles may be composed of smaller particles or their movement can be divided orbital and spin and so on. Based on quantum mechanics principles, the total quantum number M of a system is the sum of all m of orbitals and spins of all smaller particles in it, that is M=$\sum_i m_i$. Considering the above analysis, a particle (or a compound particle) with spin s=1/2 cannot be composed of two particles with spin $m_s$=1/4 because particles with $m_s$=1/4 must be far away from each other, and nor can it be composed of three particles with spin $m_s$=1/6 because such particles with the parallel spin tend to be far away from each other, and just can be composed of two particles with spin $m_s$=1/3 and $m_s$=1/6 respectively, or be composed of three particles with two spins $m_s$=1/3 and one spin $m_s$=-1/6. The electrons reported in the relevant literature can be split as a backscattered charge e/3 and a transmitted charge 2e/3,[8] which is consistent with the above analysis. In addition, the completely overlapping nature of the positive and negative phase regions of the $Y_{\frac{1}{2}, \pm\frac{1}{2}}(\theta, \varphi)$ results the fact that the Y functions of two electrons close to each other never overlap efficiently as in the atomic orbitals, so that the two electrons never merge into one. That is, there are no co-electron particle with a charge of -2e or more. In addition, the idea that the proton speculated by particle physics is composed of two up quarks with charge of 2/3e and a down quark with charge of -1/3e[9] is consistent with the analysis above, but the literature report that the spins of these quarks are s=1/2 lacks experimental evidence.

According to the above analysis, a particle with a spin of 1/3 cannot consist of two particles with spin 1/6, nor consist of a particle with a spin of 1/4 and a particle with a spin of 1/12 which have the same asymmetric spin structure and repel each other. It is only possible to consist of one particle with a spin of 1/5 and two particles with a spin of 1/15. Then there will not be particles with a spin of 1/4 in the elementary particles, and the difference between the up and down spin quantum numbers of such particles is 1/4-(-1/4)=1/2, which is a half-integer interval, and the particles with opposite spins have exactly the same spin structure, which might be the characteristics of Majorana-like particles.

Particles with spin-magnetic quantum numbers of $m_s$=1/3 can be divided into three particles with $m_s$ of 1/5, 1/15, and 1/15, that is 1/3=1/5+1/15+1/15. A particle with a magnetic quantum number of 1/5 may be significantly different from a particle with a magnetic quantum number of 1/15. In addition, the particles with $m_s$=1/5 can continue to divide, so the particle with a spin of $m_s$=1/2 can be divided as below:

$$1/2 = 1/3+1/3-1/6 \quad \text{step 1}$$
$$= (1/5+1/15+1/15) + (1/5+1/15+1/15) - (1/7+1/42) \quad \text{step 2}$$
$$= [(1/7+1/35+1/35)+1/15+1/15]*2 - (1/7+1/42) \quad \text{step 3}$$
$$=……$$

These fractional numbers may be $m_s$ quantum numbers of smaller particles or their orbital m. The angular momentum of the 3 quarks in the proton contributes only part of the proton's angular momentum [10,11] which can be explained by the analysis above. According to the decomposition method of step 2 the sum of the three main magnetic quantum numbers is $\sum m_s$=1/5+1/5-1/7=9/35, and according to step3 $\sum m_s$=1/7+1/7-1/7=1/7. The spin of proton is 1/2. The ratio of $\sum m_s$ of the three main values to 1/2 of proton spin is (9/35)/(1/2)=51.4% (~50% reported[10]) and (1/7)/(1/2)=28.6% (~30% reported[11]), respectively. This fits well with the literature data. A recent research reported that protons consist of three valence quarks, two up-type and one down-type, and gluons which hold together the three valence quarks, as well as of a 'sea' of virtual quark–antiquark pairs, which are fundamental elements of the quantum vacuum.[12] Based on this work, the paired values of m such as 1/15,1/35...... may correspond to the 'sea'.

It is indeed difficult to understand the overlap of the positive and negative phase regions of these spherical harmonics of $l$=1/n by using that of atomic orbitals. The quantum world is a challenge to our human thinking. For spin with s=$\frac{1}{n}$ or period of 2nπ, perhaps one can use a similar analogy with similar movements in the macroscopic world. For example, the moon moves around the earth, simply put, revolves around the earth every month, but in fact, the moon does not come to the starting position in one cycle, and the moon takes 12 months or 12 cycles around the earth to reach the starting position, this is like s=$\frac{1}{12}$ or period of 12×2π. The second example is a two-way closed-loop screw．The third example is that the swing head fan blades also have a rotation period of 2nπ. The quantum spin is even more unimaginable than that.

Conclusion

A series of spherical harmonic functions $Y_{l,m}(\theta,\varphi)$ with $l=\frac{1}{n}$ and $m=\pm\frac{1}{n}$ is analyzed logically based on associated Legendre equation in this paper. The values of *l* corresponding to $Y_{l,m}(\theta,\varphi)$ can also be fractional numbers such as 1/3, 1/4, … besides the spin quantum number of electrons, protons and neutrons is a fraction s = 1/2 as a quantum mechanical hypothesis. These spherical harmonics $Y_{\frac{1}{n'},\pm\frac{1}{n}}(\theta,\varphi)$ for fractional quantum numbers may provide new insights into movement of micro-particles and other scientific fields related to angular momentum. To be clear, this work is not intended to support the idea that electrons has an internal structure, it simply provides us with theory to support the possibility of angular momentum quantum numbers can be fractional and requires further experimental verification. The helical structure of the spherical harmonics $Y_{\frac{1}{n'},\pm\frac{1}{n}}(\theta,\varphi)$ results in overlap of positive and negative phase regions. Particles (or compound particles) with fractional spin could be divided into three categories according to their characteristics of $Y_{\frac{1}{n'},\pm\frac{1}{n}}(\theta,\varphi)$. The first class of particles with s=1/2, 1/6, …, like electrons, particles with the parallel spin tend to stay away from each other and particles with antiparallel spin tend to pair up. The second class of particles with s=1/3,1/5, … is not mutually exclusive whether their spin direction is the same or not. The third class of particles with s=1/4, 1/8, … will repel each other and tend to move away whether their spin direction is the same or not. The internal composition of electrons and protons is well illustrated by using fractional spin. The theory about fractional angular momentum of 1/n breaks the original theory of quantum mechanics s=1/2 and need to be further proved in theory and practice.

Acknowledgements

This work was supported by grants from educational foundations of Henan province (Grant No. HNYJS2018KC34) and Henan Normal University (Grant No. YJS2019JG02).